\documentclass[10pt,final,conference,letterpaper,romanappendices,twocolumn]{IEEEtran}
\usepackage{cite}
\usepackage{epsfig}
\usepackage{enumerate}
\usepackage{amsthm}
\usepackage{amsmath,amssymb,amsfonts,amstext,amsbsy,amsopn,dsfont}
\usepackage{cases}
\usepackage{sublabel}
\usepackage{array}

\newtheorem{theorem}{\textbf{Theorem}}

\newtheorem{remark}{\textbf{Remark}}

\newcommand{\defn}{\triangleq}
\newcommand{\dif}{\textmd{d}}
\newcommand{\subB}{\mathsf{B}}
\newcommand{\subU}{\mathsf{U}}

\begin{document}

\title{Random Cell Association and Void Probability in Poisson-Distributed Cellular Networks}

\author{Chun-Hung Liu and Li-Chun Wang \\Department of Electrical and Computer Engineering \\National Chiao Tung University, Hsinchu, Taiwan\\ e-mail: chungliu@nctu.edu.tw and lichun@g2.nctu.edu.tw}

\maketitle

\begin{abstract}
 This paper studied the fundamental modeling defect existing in Poisson-distributed cellular networks in which all base stations form a homogeneous Poisson point process (PPP) of intensity $\lambda_{\subB}$ and all users form another independent PPP of intensity $\lambda_{\subU}$. The modeling defect, hardly discovered in prior works, is the void cell issue that stems from the independence between the distributions of users and BSs and ``user-centric'' cell association, and it could give rise to very inaccurate analytical results. We showed that the void probability of a cell under generalized random cell association is always bounded above zero and its theoretical lower bound is $\exp\left(-\frac{\lambda_{\subU}}{\lambda_{\subB}}\right)$ that can be achieved by large association weighting. An accurate expression of the void probability of a cell was derived and simulation results validated its correctness. We also showed that the associated BSs are essentially no longer a PPP such that modeling them as a PPP to facilitate the analysis of interference-related performance metrics may detach from reality if the BS intensity is not significantly large if compared with the user intensity.    
\end{abstract}

\section{Introduction}
 Due to the stunning increase of mobile data traffic in cellular networks,  deploying more and different types of base stations (BSs) in cellular networks tends to be an effective means of boosting network coverage and capacity. In recent years,  modeling and analysis in heterogeneous cellular networks (HetNets) have gained lots of research attention since  traditional and simple cellular modeling and analysis cannot correctly and accurately characterize the distribution feature of HetNets \cite{ADWYNER94,JXJZJGA11,PXCHLJGA13}. Stochastic geometry has been shown to be a powerful mathematical tool for modeling the random distribution of BSs since it makes the mathematical analysis of some network performance metrics, such as coverage probability and average rate, much more tractable, and it even leads to closed-form results in some cases \cite{MHJGAFBODMF09}\cite{JGAFBRKG11}.           
 
 In the framework of stochastic geometry for cellular networks, users are modeled as an independent Poisson point process (PPP) in addition to the PPPs formed by all BSs \cite{JGAFBRKG11}\cite{HSDRKGFBJGA12}.  Usually, ``user-centric'' cell association is adopted in the network, i.e., every user tries to associate with its best service BS via some cell association schemes. For example, if all users are looking for a BS that is able to provide the long-term strongest signal power, they will associate with their nearest BS if they completely eliminate fading and/or shadowing effects on their channels.  Such nearest cell association is a popular scheme frequently adopted in many prior works since the distribution of the distance between a user and its associated BS can be easily found, and most importantly the cells of all macro BSs that consist of the entire network plane can be characterized by Voronoi tessellation and thus the average number of users in a cell can be accurately estimated by either simulated or theoretical computation if the cell of a BS is Voronoi-tessellated \cite{DSWKJM96}\cite{FBBBL10}.  
 
  \begin{figure}
  	\centering
  	\includegraphics[width=3.5in, height=2.5in]{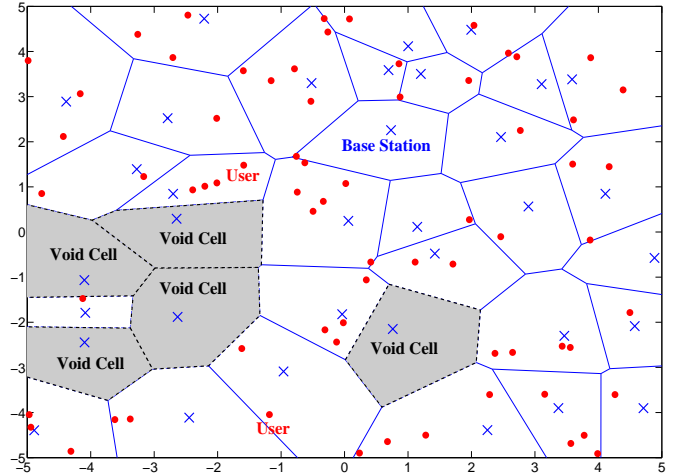}
  	\caption{An illustration example of void cells in a cellular network. BSs and users are two independent PPPs and the cells of BSs are created by Voronoi tessellation. The intensity ratio of users to base stations is 2.}
  	\label{Fig:VoidCell}
  \end{figure}
 
 \subsection{Motivation and Prior Work}
 User-centric cell association has an intrinsic defect that gives rise to an unreasonable modeling phenomenon in Poisson-distributed cellular networks. Namely, it could make BSs not associated by any users in the network, which seldom happens in reality especially for macro BSs.  This void cell (BS) phenomenon is illustrated in Fig. \ref{Fig:VoidCell} for the case of Voronoi-tessellated cells. Users can be imaged to be dropped on the plane consisting of the cells so that void cells very likely exist in the network if the user intensity is not significantly larger than the BS intensity. For example, consider there are $n$ users uniformly distributed in a unit-area network which is tessellated by $n$ equal-sized cells of BSs, and thus the probability of having a void cell is $\left(1-\frac{1}{n}\right)^n$ that is close to $e^{-1}\approx 0.367$ as $n$ is very large. This shows that the void cell issue cannot be ignored in a cellular network whose BS intensity is more or less equal to the user intensity, such as hyper-dense small cell networks.
 
 Despite the important phenomenon of void cells, there are very few works that study and model the effect of void cells, and almost all prior works on cell association in Poisson cellular networks overlook this problem \cite{JGAFBRKG11,HSDRKGFBJGA12,HSJYJSXPJGA12,PXCHLJGA13,MDRAGGEC13,SSJGA14}. Without modeling the effect of void cells, the analyses of network performance metrics, e.g. coverage probability and average rate, are certainly underestimated since the void BSs do not contribute any interference. Although references \cite{SLKH12}\cite{CLJZKBL14} do consider the void cell impact on their models, the void probability of a cell they found is only valid for nearest cell association, and the associated BSs are still viewed as a homogeneous PPP in these two works, which is certainly questionable due to the correlation between the associated BSs. 
 
 \subsection{Contributions}
 In this paper, our goal is to delve the fundamental correlation between cell association and void cell probability. To make the model simple, only a single tier of BSs that form a homogeneous PPP of intensity $\lambda_{\subB}$ is considered and users form another independent PPP of intensity $\lambda_{\subU}$. A random cell association (RCA) scheme was proposed to cover several cell association schemes, such as nearest cell association, strongest received power association, etc.. Our first contribution is to theoretically show that the achievable lower bound on the void probability of a cell is $\exp\left(-\frac{\lambda_{\subU}}{\lambda_{\subB}}\right)$. An accurate closed-form result of the void probability was derived under the RCA scheme for a more practical channel model that characterizes path loss as well as the composite effect of Nakagami-$m$ fading and log-normal shadowing. According to the derived void probability, we can gain some insights on how the void probability is affected by channel impairments and how its theoretical lower bound is achieved by RCA. Finally, we discovered and numerically verified that the associated BSs are essentially not a PPP and modeling them as a PPP could give rise to fairly inaccurate analysis especially in HetNets and small cell networks deployed with high BS intensity.  
 
\section{System Model and Preliminaries}\label{Sec:SysModDefns}
%\subsection{The PPP-based model of a $K$-tier HetNet}
Consider an infinitely large wireless cellular network on the plane $\mathbb{R}^2$ in which all base stations form a marked homogeneous PPP $\Phi_{\subB}$ of intensity $\lambda_{\subB}$ given by
\begin{align}\label{Eqn:KtierModelBS}
\Phi_{\subB} \defn &\bigg\{(B_{i},V_{i}, H_{i}, \mathcal{C}_{i}): B_{i}\in\mathbb{R}^2, V_{i}\in\{0,1\}, H_{i}\in\mathbb{R}_+, \nonumber \\
 &\mathcal{C}_{i}\subset\mathbb{R}^2, \forall i\in\mathbb{N}_{++} \bigg\},
\end{align}
where $B_{i}$ denotes the $i$th nearest BS in $\Phi_{\subB}$ to the origin and its location, the cell of $B_i$ is represented by $\mathcal{C}_i$ and it contains all \textit{potential} users serviced by BS $B_i$\footnote{The cell of a base station and its statistical properties can be characterized once the cell tessellation algorithm (e.g., Voronoi tessellation for a PPP) over the plane is designated.},   $V_i$ is a void cell index that indicates whether or not at least one user is in the cell of $B_i$, i.e., whether $\mathcal{C}_i\cap\Phi_{\subU}=\emptyset$ is true or not -- $V_i$ is equal to one if $\mathcal{C}_i\cap\Phi_{\subU}\neq\emptyset$, otherwise zero, $H_i$ is used to characterize the downlink channel power gain from $B_i$ to its service user. In addition, throughout this paper we assume all cells are disjoint, i.e. $\mathcal{C}_i\cap\mathcal{C}_j=\emptyset$ for all $i\neq j$ and $ \bigcup_{i=1}^{\infty} \mathcal{C}_i\subseteq\mathbb{R}^2$. Transmit powers of all BSs are the same  and all users form another independent homogeneous Poisson point process $\Phi_{\subU}$ of intensity $\lambda_{\subU}$.   

To characterize the statistical behavior of downlink channel power gains in a much general form, we assume that all $H_i$'s are i.i.d. random variables whose probability density function (pdf) characterizes the composite effect of Nakagami-$m$ fading and log-normal shadowing and is given as follows \cite{STUBER11}
\begin{align}\label{Eqn:PdfChannelGain}
f_{H}(h) =& \frac{m^mh^{m-1}}{\Gamma(m)\sqrt{2\pi \sigma^2}} \int_{0^+}^{\infty} x^{-(m+1)} \exp\bigg(-\frac{mh}{x}\nonumber\\
&-\frac{(\ln x-\mu)^2}{2\sigma^2}\bigg) \dif x,  
\end{align} 
where $\Gamma(x)=\int_{0}^{\infty} t^{x-1}e^{-t}\dif t$ is the gamma function, $\mu$ and $\sigma^2$ are the mean and variance of log-normal shadowing, respectively. Thus, $H_i$ is essentially a gamma-log-normal random variable. Note that the mean of $H_i$ is $\exp\left(\mu+\frac{\sigma^2}{2}\right)$. Without loss of generality, the following downlink analysis will be based on a typical (reference) user $U_0$ located at the origin. Each user associates with a base station in $\Phi_{\subB}$ by using random cell association (RCA).  In a given $\Phi_{\subB}$, the RCA scheme associates user $U_0$ with base station $B^*_0$  by the following rule
\begin{equation}\label{Eqn:RCA}
B_0^{*}=\arg\sup_{B_i\in\Phi_{\subB}} \left(W_iH_i\|B_i\|^{-\alpha}\right),
\end{equation}
where $W_i$'s are the i.i.d. random weightings for BS association and $\|B_i\|$ denotes the Euclidean distance between $B_i$ and the origin. RCA can cover several BS association schemes. For example, if the channel state information $H_i$ is available $W_i$ can be assigned as $W_i=H_i^{-1}$, RCA reduces to the popular \textit{nearest} BS association.  If all the weightings of RCA are the same constant, RCA is essentially the association scheme of strongest received power, which is suitable for non-stationary users that cannot acquire the mean received power, e.g., they are moving very fast. RCA is a user-centric scheme, i.e., it is able to make every user associate with certain base station. In other words, no users are blocked out of the network.  

Although user-centric cell association ensures no blocked users in the network, it cannot, as we will show later, guarantee that every BS is associated with at least one user, i.e., the probability that a cell/BS is not associated with any users, called the void probability of a cell, is always bounded above zero. Since users are a homogeneous PPP of intensity $\lambda_{\subU}$, the void probability of a cell in the network can be written as
\begin{equation}
p_{\emptyset} = \mathbb{P}[V_i=0]=\int_{\mathbb{R}_+} e^{-\lambda_{\subU}x}f_{\nu(\mathcal{C}_i)}(x) \dif x,
\end{equation}  
where $\nu(\mathcal{A})$ is the Lebesgue measure of the bounded Borel set $\mathcal{A}\subset\mathbb{R}^2$ and $f_{Z}(z)$ is the pdf of random variable $Z$. In order to explicitly calculate $p_{\emptyset}$, one has to know $f_{\nu(\mathcal{C}_i)}(x)$ for the RCA scheme. The pdf of the cell area of a BS depends on the adopted cell association scheme and can be accurately characterized by the conservation property of a homogeneous PPP introduced in the following subsection. The accurate pdf of the cell for RCA will be derived in Section \ref{Sec:CellVoidProb}. 

\subsection{Random Conservation Property of a PPP}
In this subsection, we introduce the random conservation property of a PPP, which specifies how the intensity of a PPP is changed after all points of the PPP are transformed by i.i.d. random mapping matrices. The random conservation property is specified in the following theorem.
\begin{theorem}[Random Conservation Property of a PPP]\label{Thm:RanTransProp}
Suppose $\hat{\Phi}$ is an independently marked PPP of intensity measure $\hat{\Lambda}$ on $\mathbb{R}^{\mathrm{d}'}$, which can be expressed as follows
\begin{equation}
\hat{\Phi}\defn\{(X_i, \mathbf{T}_i): X_i\in\mathbb{R}^{\mathrm{d}'}, \mathbf{T}_i\in\mathbb{R}^{\mathrm{d}\times \mathrm{d}'}, \forall i\in\mathbb{N}_{++}\},
\end{equation}
where $X_i$ denotes node $i$ and its location and $\mathbf{T}_i: \mathbb{R}^{\mathrm{d}'}\rightarrow\mathbb{R}^{d}$ is the non-singular mapping matrix (operator) of node $X_i$. For all $i\neq j$, $\mathbf{T}_i$ and $\mathbf{T}_j$ are (element-wise) i.i.d.. Let $\hat{\Phi}^{\dag}$ be the mapped point process on $\mathbb{R}^{\mathrm{d}}$ generated by using the random mapping matrix of each point in $\hat{\Phi}$, i.e., it is defined as
\begin{equation}
\hat{\Phi}^{\dag}\defn\{X_i^{\dag}\defn\mathbf{T}_i (X_i): X_i\in\hat{\Phi}, \mathbf{T}_i\in\mathbb{R}^{\mathrm{d}\times \mathrm{d}'}, \forall i\in\mathbb{N}_{++}\}.
\end{equation}
Then $\hat{\Phi}^{\dag}$ is also a PPP and for any bounded Borel set $\mathcal{A}\subset \mathbb{R}^{\mathrm{d}}$ its intensity measure $\hat{\Lambda}^{\dag}$ is given by
\begin{equation}\label{Eqn:TransGenIntensityMeasure}
\hat{\Lambda}^{\dag}(\mathcal{A}) = \hat{\Lambda}(\mathcal{A}')\mathbb{E}\left[\frac{1 }{\sqrt{\det\left(\mathbf{T}^{\mathrm{T}}\mathbf{T}\right)}}\right]
\end{equation}
if $\nu_{\mathrm{d}'}(\mathcal{A}')=\nu_{\mathrm{d}}(\mathcal{A})$, where $\mathcal{A}'\subset\mathbb{R}^{\mathrm{d}'}$ is also a bounded Borel set and $\mathbf{T}^{\mathrm{T}}$ is the transpose of $\mathbf{T}$. If $\hat{\Phi}$ is homogeneous and has an intensity $\hat{\lambda}$, $\hat{\Phi}^{\dag}$ is also homogeneous and has the following intensity 
\begin{equation}\label{Eqn:TransHomoIntensityMeasure}
\hat{\lambda}^{\dag} = \hat{\lambda} \mathbb{E}\left[\frac{1}{\sqrt{\det(\mathbf{T}^{\mathrm{T}}\mathbf{T}) }}\right].
\end{equation}
\end{theorem}
\begin{IEEEproof}
See Appendix \ref{App:ProofRanTransProp}.
\end{IEEEproof}
\begin{remark}
Theorem \ref{Thm:RanTransProp} is a generalization of the conservation property in \cite{DSWKJM96}\cite{CHLBRSC15}. In a special case of $\mathrm{d}=\mathrm{d}'=2$, all points in $\hat{\Phi}^{\dag}$ are mapped from their corresponding points in a homogeneous PPP $\hat{\Phi}$ by scaling them with random variables $T$. In this case, $\mathbf{T}_i$ is equal to $\text{diag}(T, T)$ and thus $\hat{\lambda}^{\dag}$ is equal to $\hat{\lambda} \mathbb{E}[1/T^{2}]$.  
\end{remark}

The random conservation property can significantly reduce the complexity of analyzing the statistics of some performance metrics induced by a PPP, especially a homogeneous PPP with i.i.d. marks. For instance, the RCA scheme in \eqref{Eqn:RCA} can be further simplified as
\begin{equation}
\|B_0^*\|\stackrel{d}{=}(WH)^{\frac{1}{\alpha}} \left(\inf_{B^{\dag}_i\in\Phi^{\dag}_{\subB}} \|B^{\dag}_i\|\right),
\end{equation}
where $\stackrel{d}{=}$ stands for equivalence in distribution,  $B^{\dag}_i\defn (W_iH_i)^{-\frac{1}{\alpha}}B_i$, $\Phi^{\dag}_{\subB}$ is a homogeneous PPP of intensity $\lambda^{\dag}_{\subB}=\lambda_{\subB}\mathbb{E}\left[(WH)^{\frac{2}{\alpha}}\right]$ based on \eqref{Eqn:TransHomoIntensityMeasure} in Theorem \ref{Thm:RanTransProp}. Hence, for the typical user $U_0$ with RCA, its serving distance can be statistically  instead found by using the transformed new PPP $\Phi^{\dag}_{\subB}$. In other words, the random conservation property can make RCA reduce to nearest BS association, which significantly simplifies the analysis of the performance metric of RCA, such as the coverage/outage probability of RCA, since many existing results of nearest BS association can be applied in this context by simply modifying them with an updated intensity of the BSs.

\section{Void Probability of a Cell under RCA}\label{Sec:CellVoidProb}
As pointed out in the previous section, any ``user-centric'' association scheme cannot guarantee that there is at least one user in each cell, i.e., a void cell could exist in the network. This phenomenon can be intuitively interpreted by using a Poisson-Dirichlet (Voronoi) tessellation for a PPP of BSs. Suppose the Voronoi tessellation is used to determine the cell of each BS in $\Phi_{\subB}$ and all users adopt the nearest BS association scheme to connect with their serving BS. This nearest associating process can be viewed as the process of dropping all users in $\Phi_{\subU}$ on the large plane consisting of the Voronoi-tessellated cells formed by $\Phi_{\subB}$. Under this circumstance, the probability mass function (pmf) of the number of users in a cell of $\Phi_{\subB}$ can be expressed as 
\begin{equation}\label{Eqn:PmfNumUserVorCell}
p_n\defn\mathbb{P}[\Phi_{\subU}(\mathcal{C})=n] = \mathbb{E}\left[\frac{(\lambda_{\subU}\nu(\mathcal{C}))^n}{n!}e^{-\lambda_{\subU}\nu(\mathcal{C})}\right],
\end{equation}
where $\mathcal{C}$ denotes a Voronoi cell of a BS in $\Phi_{\subB}$, $\Phi_{\subU}(\mathcal{C})$ represents either the user set in cell $\mathcal{C}$  or the number of users in cell $\mathcal{C}$, and $\nu(\mathcal{C})$ is the Lebesgue measure of $\mathcal{C}$.

Unfortunately, the theoretical result of $p_n$ in \eqref{Eqn:PmfNumUserVorCell} is unknown since the pdf of a Voronoi cell area is still an open problem \cite{DSWKJM96}. However, it can be accurately approximated by using a Gamma distribution with some particular parameters \cite{DSWKJM96,JSFNZ07}. Reference \cite{JSFNZ07} suggests the following Gamma distribution for $f_{\nu(\mathcal{C})}(x)$:
\begin{equation}\label{Eqn:ApporxPdfVoronoiCell}
f_{\nu(\mathcal{C})}(x) = \frac{(\zeta \lambda_{\subB}x)^{\zeta}}{\Gamma(\zeta)x}e^{-\zeta\lambda_{\subB}x}, 
\end{equation} 
where $\Gamma(x)=\int_{0}^{\infty} t^{x-1}e^{-t}\dif t$ is the Gamma function and $\zeta=\frac{7}{2}$ can achieve an accurate pdf of a Voronoi cell area. Substituting \eqref{Eqn:ApporxPdfVoronoiCell} into  \eqref{Eqn:PmfNumUserVorCell} yields the following result:   
\begin{align}
p_n
&=\frac{\lambda^n_{\subU}}{n!}\frac{(\zeta \lambda_{\subB})^{\zeta}}{\Gamma(\zeta)}\int_{0}^{\infty}   x^{n+\zeta-1}  e^{-(\zeta\lambda_{\subB}+\lambda_{\subU})x}\dif x \nonumber\\
&=\frac{1}{n!}\frac{\Gamma(n+\zeta)}{\Gamma(\zeta)} \frac{\lambda^n_{\subU}(\zeta \lambda_{\subB})^{\zeta}}{(\zeta\lambda_{\subB}+\lambda_{\subU})^{(n+\zeta)}}. \label{Eqn:AppPmfNumUserVorCell}
\end{align} 
Hence, the pmf of the number of users in a cell for nearest BS association has a closed-form expression. The void probability of a cell, $p_{\emptyset}$, can be found by $p_n$ with  the case of $n=0$. Namely,
\begin{equation}\label{Eqn:VoidProbNearBS}
p_{\emptyset}=\left(1+\frac{\lambda_{\subU}}{\zeta\lambda_{\subB}}\right)^{-\zeta}
\end{equation}
and this indicates that the intensity of the non-associated BSs is $\lambda_{\subB}p_{\emptyset}$ that is not negligible especially when $\lambda_{\subU}/\lambda_{\subB}$ is small. Most importantly, \textit{the results in \eqref{Eqn:ApporxPdfVoronoiCell} and \eqref{Eqn:AppPmfNumUserVorCell} are no longer accurate for all non-nearest BS association schemes since users do not necessarily associate with their nearest BS, such as the RCA scheme mentioned in Section \ref{Sec:SysModDefns}}. However, an accurate void probability of a BS under the RCA scheme can be derived as shown in the following. 

Although the pdf of a Voronori cell is unknown, its mean can be shown to be $1/\lambda$ if the Voronori cell is created by a homogeneous PPP of intensity $\lambda$ \cite{DSWKJM96}. As a result,  the lower bound on $p_n$ in \eqref{Eqn:PmfNumUserVorCell} for $n=0$ is given by
\begin{equation}
p_{\emptyset}=\mathbb{E}\left[e^{-\lambda_{\subU}\nu(\mathcal{C})}\right] \geq \exp\left(-\frac{\lambda_{\subU}}{\lambda_{\subB}}\right)
\end{equation}
due to Jensen's inequality.  This lower bound on the void probability reveals three crucial implications: (i) the void probability of a cell is always bounded above zero  such that there could be a certain number of void BSs in the cellular network, (ii) nearest BS association cannot achieve this lower bound since its void probability in \eqref{Eqn:VoidProbNearBS} is always larger than the lower bound. (iii) From an energy-saving point of view, the lower bound represents the maximum percentage of all consumed energy in the network that can be saved. Later, we will theoretically show that, \textit{this lower bound can be achieved by RCA}. 

To derive an accurate void probability of a cell under the RCA scheme, we approach this problem from a fundamental connectivity point of view and derive the bounds on $p_{\emptyset}$ as shown in the following theorem.

 \begin{theorem}\label{Thm:BoundVoidProb}
If all users in $\Phi_{\subU}$ adopt the RCA scheme defined in \eqref{Eqn:RCA} to associate with a BS $\Phi_{\subB}$ defined in \eqref{Eqn:KtierModelBS},  the bounds on the void probability of a cell in $\Phi_{\subB}$ are given by
\begin{equation}\label{Eqn:BoundVoidProb}
\left(1+\frac{\lambda_{\subU}}{\zeta^{\dag}\lambda_{\subB}}\right)^{-\zeta^{\dag}}\geq p_{\emptyset}\geq \exp\left(-\frac{\lambda_{\subU}}{\lambda_{\subB}}\right),
\end{equation} 
where  $\zeta^{\dag}\defn\mathbb{E}\left[(WH)^{\frac{2}{\alpha}}\right]\mathbb{E}\left[(WH)^{-\frac{2}{\alpha}}\right]$.
 \end{theorem} 
 \begin{IEEEproof}
 See Appendix \ref{App:ProofBoundVoidProb}.
 \end{IEEEproof}
  \begin{remark}\label{Rem:VoidBSsNotPPP}
   According to the proof of Theorem \ref{Thm:BoundVoidProb}, the association events of different BSs could be correlated such that the resulting process of associated BSs is no longer a PPP even though its intensity is $(1-p_{\emptyset})\lambda_{\subB}$.
   \end{remark}
 The lower bound on $p_{\emptyset}$ is derived while considering the complete independence exists between the non-associated events of a BS, whereas the upper bound is obtained by approaching the opposite case, i.e., all non-associated events of a BS are caused by the farthest user in its cell and thus they are highly correlated. Accordingly, it is reasonable to conjecture that the upper bound is tightly close to the lower bound provided that the cross-correlations between all non-associated events are significantly weakened.  On the other hand, mathematically we have the following
 \begin{equation}
 \lim_{\zeta^{\dag}\rightarrow\infty} p_{\emptyset}=\exp\left(-\frac{\lambda_{\subU}}{\lambda_{\subB}}\right)
 \end{equation}
 since the upper bound approaches to the lower bound $\exp(-\lambda_{\subU}/\lambda_{\subB})$ as $\zeta^{\dag}$ goes to infinity and thus the bounds in \eqref{Eqn:BoundVoidProb} are very tight as $\zeta^{\dag}$ becomes large. This intuitively reveals that large $\zeta^{\dag}$ will  ``de-correlate'' all non-associated events, which is an important observation since it indicates that the theoretical minimum value of $p_{\emptyset}$ is $\exp\left(-\frac{\lambda_{\subU}}{\lambda_{\subB}}\right)$ and it can be achieved by the large $\frac{2}{\alpha}$-fractional moment of $WH$. For example, if $W=1$, $p_{\emptyset}=\exp\left(-\frac{\lambda_{\subU}}{\lambda_{\subB}}\right)$ can be achieved if channels have high shadowing power. This implies that the void probability is reduced when more users join the network under a given BS intensity. In other words, when the network has a large user population the efficacy of reducing $p_{\emptyset}$ by using large $\zeta^{\dag}$ is apparently undermined such that the performance of RCA is similar to that of nearest BS association in this case. 
 
 Although the bounds on $p_{\emptyset}$ are characterized, an accurate result of $p_{\emptyset}$ is still needed since it can help us understand how many cells per unit area are void. The following theorem renders an accurate closed-form expression of $p_{\emptyset}$ under RCA.  
  \begin{theorem}\label{Thm:ExacVoidProb}
  If RCA is adopted, the void probability of a cell can be accurately characterized by
 \begin{equation}\label{Eqn:VoidProbRCAwAlpha}
  p_{\emptyset}=\left(1+\frac{\lambda_{\subU}}{\rho\lambda_{\subB}}\right)^{-\rho},
 \end{equation}
 where $\rho=\frac{7}{2}\mathbb{E}\left[(WH)^{\frac{2}{\alpha}}\right]\mathbb{E}\left[(WH)^{-\frac{2}{\alpha}}\right]=\frac{7}{2}\zeta^{\dag}$.
  \end{theorem}
  \begin{IEEEproof}
  Since $\left(1+x/b\right)^{-b}\geq\left(1+x/a\right)^{-a}$ for $a>b$ and  $\rho>\zeta^{\dag}$, we know 
  $$\left(1+\frac{\lambda_{\subU}}{\lambda_{\subB}\zeta^{\dag}}\right)^{-\zeta^{\dag}}>\left(1+\frac{\lambda_{\subU}}{\rho\lambda_{\subB}}\right)^{-\rho}$$
  and thus $p_{\emptyset}$ in \eqref{Eqn:VoidProbRCAwAlpha} is between the bounds in \eqref{Eqn:BoundVoidProb}. Also, for nearest BS association the void probability of a BS with $\zeta=\frac{7}{2}$ in \eqref{Eqn:VoidProbNearBS} is very accurate. Therefore, we can conclude that $p_{\emptyset}$ with $\rho=\frac{7}{2}\zeta^{\dag}$ is an accurate void probability of a BS for the RCA scheme since such $\rho$ reduces to $\frac{7}{2}$ as RCA reduces to nearest BS association (i.e., $W_i=H_i^{-1}$ for all $i$). 
  \end{IEEEproof}
  
 According to Remark \ref{Rem:VoidBSsNotPPP} and Theorem \ref{Thm:ExacVoidProb}, a couple of  important observations can be summarized as follows. 
  \begin{itemize}
  	\item For a cellular network with the condition $\frac{\lambda_{\subU}}{\lambda_{\subB}}\gg 1$, modeling the associated BSs as  a homogeneous PPP of intensity $(1-p_{\emptyset})\lambda_{\subB}$  is still acceptable since it just slightly looses accuracy in analysis. Hence, a cellular network with only macro BSs is hardly affected by the void cell problem, but (hyper-dense) small cell and heterogeneous networks usually would suffer the void cell problem.  
  	\item  Large $\zeta^{\dag}$ can considerably reduce the void probability of a cell. In other words, if the association weighting $W_i$ is designed such that users favor a BS with large channel power the void cell problem can be quietly mitigated. 
  \end{itemize}
  
\section{Simulation Results}
\begin{figure}
  \centering
  \includegraphics[width=3.75in, height=2.5in]{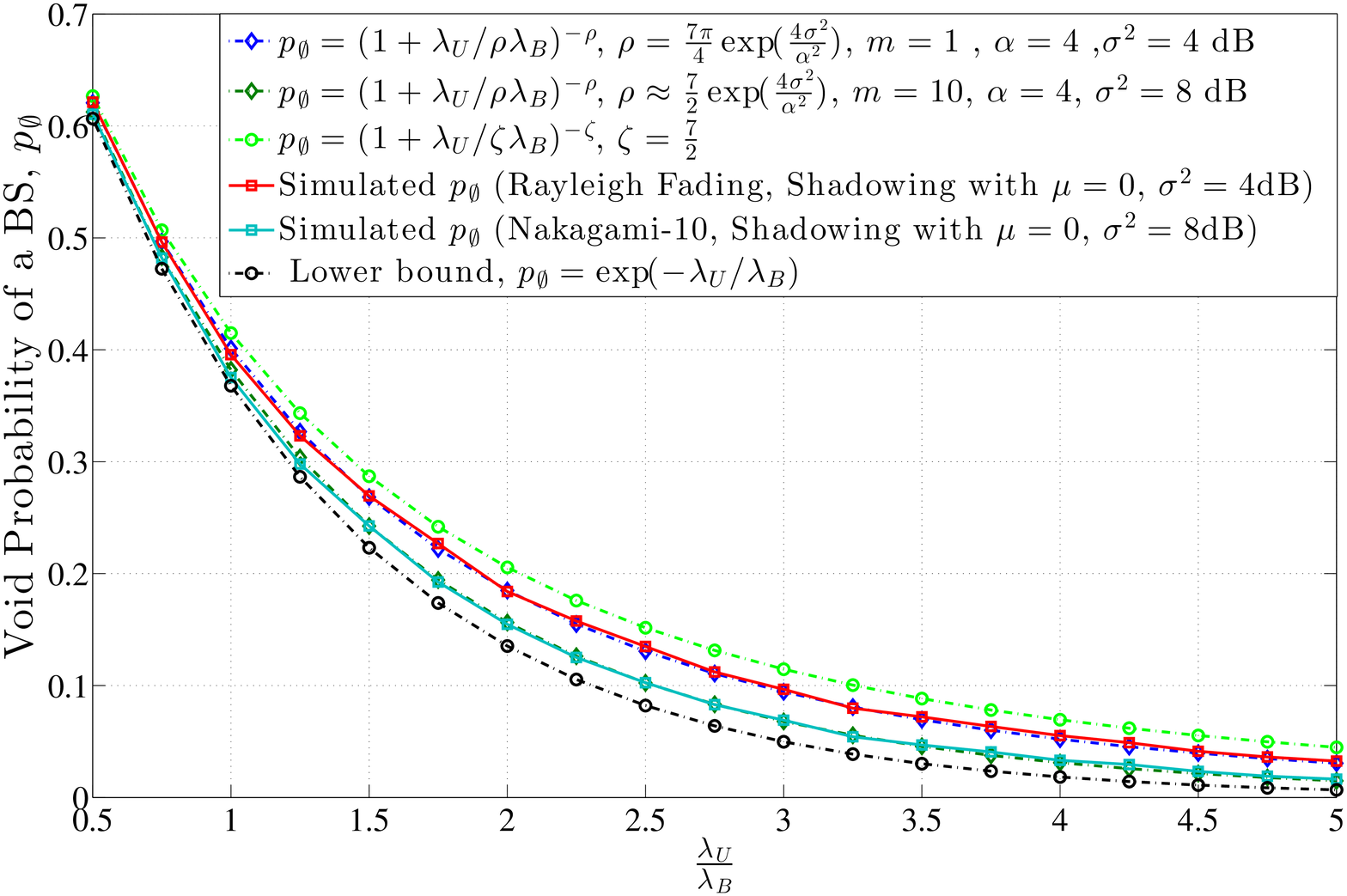}
  \caption{The void probability of a cell for the two cases of RCA: nearest BS association ($W_i=H^{-1}_i$) and strongest received power association $W_i=1$. The path loss exponent $\alpha$ for simulation is $4$ and $\lambda_{\subU}=370$ users/km$^2$.} 
  \label{Fig:VoidProbCell}
  \end{figure}

The accurate result of the void probability of a cell has been shown in \eqref{Eqn:VoidProbRCAwAlpha}. To verify its correctness, we simulate it for the two cases of RCA, i.e., nearest BS association ($W_i=H^{-1}_i$) and strongest received power association ($W_i=1)$ for all $i$. For nearest BS association, $\rho$ in \eqref{Eqn:VoidProbRCAwAlpha} is $\frac{7}{2}$, while for strongest received power association we have
\begin{equation}\label{Eqn:RhoStrRecPowAss}
\rho = \frac{7}{2}\mathbb{E}\left[H^{\frac{2}{\alpha}}\right]\mathbb{E}\left[H^{-\frac{2}{\alpha}}\right]=\frac{7\Gamma(m+\frac{2}{\alpha})\Gamma(m-\frac{2}{\alpha})}{2[\Gamma(m)]^2}e^{\left(\frac{4\sigma^2}{\alpha^2}\right)}.
\end{equation}  
Fig. \ref{Fig:VoidProbCell} shows the simulation results of the void probability for the two cell association  cases. As expected, the void probability of the nearest BS association scheme is no longer accurate if users associate with their BS with the strongest received power.  For example, when $\frac{\lambda_{\subU}}{\lambda_{\subB}}\approx 2$, the void probabilities for the nearest BS association and strongest received power association schemes with Rayleigh fading and $8$-dB shadowing are $0.2$ and $0.14$, respectively. The lower bound on the void probability is around 0.135. The void probability given in \eqref{Eqn:VoidProbRCAwAlpha} indeed accurately coincides with the simulated result and is much closer to the lower bound. Also, simulation results and \eqref{Eqn:RhoStrRecPowAss} both indicate that \textit{shadowing can significantly reduce the void probability whereas fading just slightly reduces it}.  
 
\begin{figure}[t!]
 	\centering
 	\includegraphics[width=3.65in, height=2.25in]{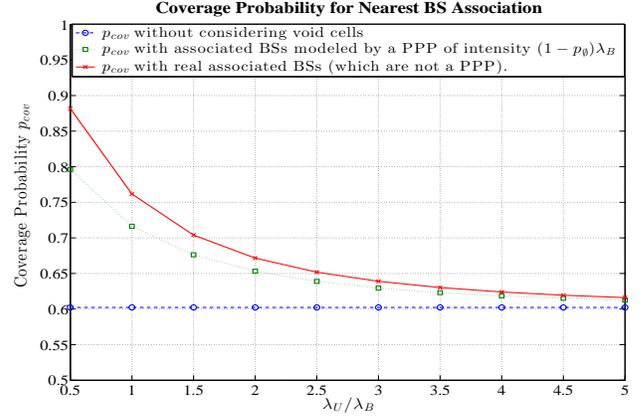}
 	\caption{The coverage probability of a user for  the nearest BS association scheme ($W_i=1/H_i$). The network parameters for simulation are $\alpha=4$,  $\mu=0$, $\sigma^2=4$dB, $\beta=0.8$, $m=1$ and $\lambda_{\subU}=370$ users/km$^2$.} 
 	\label{Fig:CovProbNearBSAss}
 \end{figure}

\begin{figure}[t!]
  	\centering
  	\includegraphics[width=3.65in, height=2.25in]{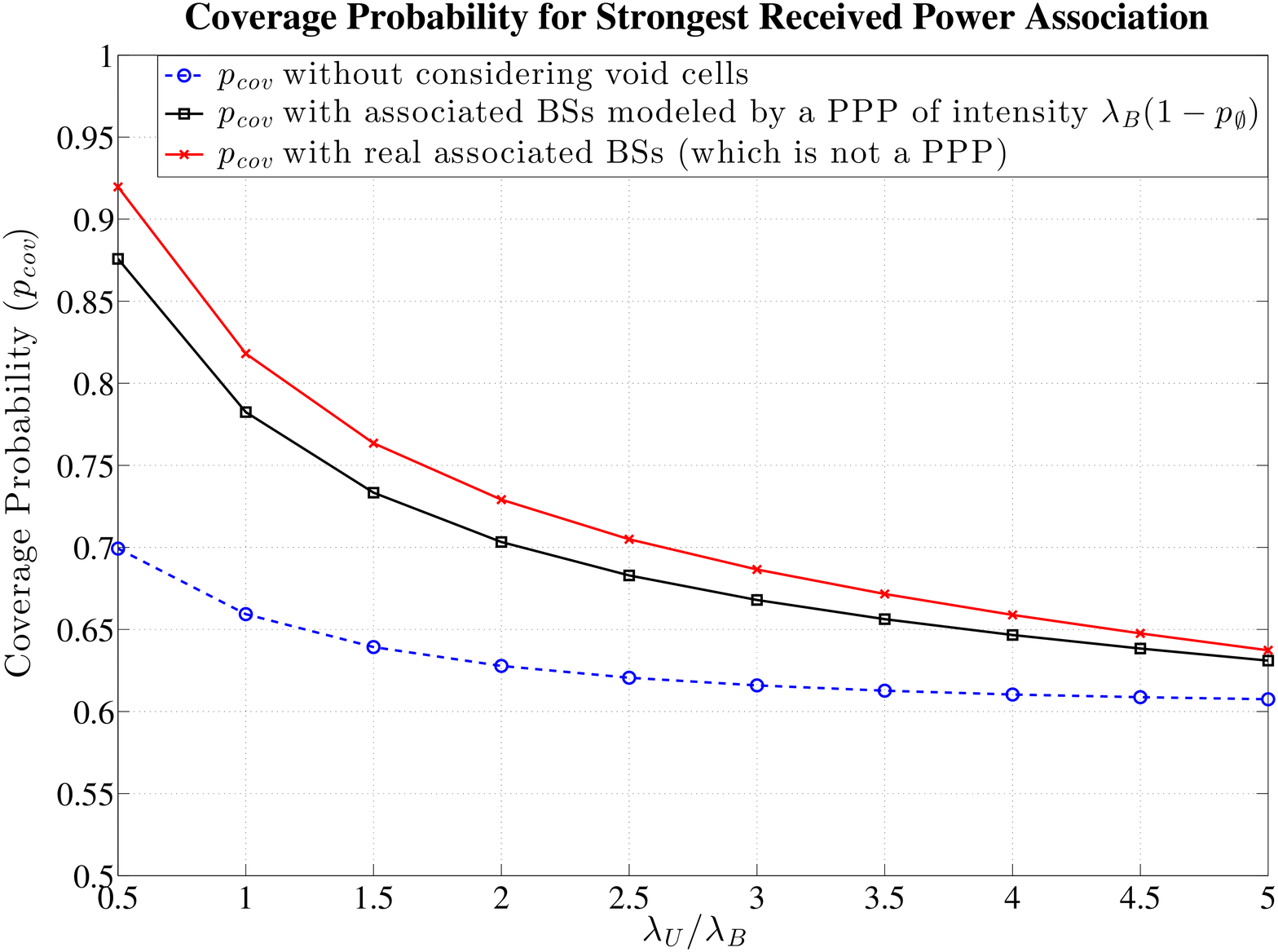}
  	\caption{The coverage probability of a user for  the strongest received power association scheme ($W_i=1$). The network parameters for simulation are $\alpha=4$,  $\mu=0$, $\sigma^2=4$dB, $\beta=0.8$, $m=1$ and $\lambda_{\subU}=370$ users/km$^2$.} 
  	\label{Fig:CovProbStrPowAss}
  \end{figure}

 Next, we simulate how RCA and the void cell problem impact the coverage probability of a user. The coverage probability of the typical user $U_0$ can be written as
 \begin{equation}
 p_{cov} = \mathbb{P}\left[\text{SIR}_0\geq \beta\right],
 \end{equation}
 where $\text{SIR}_0$ is the signal-to-interference ratio (SIR) evaluated at $U_0$ and $\beta$ is the SIR threshold for successful decoding. The simulation results for nearest BS and strongest received power association schemes are shown in Figs. \ref{Fig:CovProbNearBSAss} and \ref{Fig:CovProbStrPowAss}, respectively. As shown in the two figures, the case without considering the void cell problem seriously underestimates the real coverage probability in the region of $\frac{\lambda_{\subU}}{\lambda_{\subB}}\leq 1$.  Modeling the associated BSs as a PPP of intensity $(1-p_{\emptyset})\lambda_{\subB}$ can attain a more accurate coverage probability, but still not very accurate in the region of $\frac{\lambda_{\subU}}{\lambda_{\subB}}\leq 1$. Note that the coverage probability without considering the void cell problem in Fig. \ref{Fig:CovProbNearBSAss} does not depend on the BS intensity, which has been claimed in \cite{JGAFBRKG11}. However, in fact the coverage probability is indeed affected by the BS intensity since the void probability issue is counted in the model and it is a function of the BS intensity.  
  
\section{Concluding Remarks}
The fundamental relationship between random cell association and void cell probability is exploited in this paper, which is an important issue yet overlooked in almost all prior works.  Random cell association that can cover several cell association schemes was proposed, and the accurate result of the void probability of a cell is derived and its correctness was verified by simulation. Both theoretical and simulation results suggest that the void cell problem should be considered and properly deal with for cellular networks with small $\lambda_{\subU}/\lambda_{\subB}$. In order to easily characterize the void probability under RCA, only one tier of BSs is considered in this paper. Our future work will consider an HetNet with more tiers of BSs and focus on finding the accurate closed-form result of the void probability of a cell in any tier of the HetNet.

\appendix
\subsection{Proof of Theorem \ref{Thm:RanTransProp}}\label{App:ProofRanTransProp}
The following proof essentially generalizes the proof of Lemma 1 in \cite{CHLBRSC15}. First, we define the void probability of $\hat{\Phi}\cap\mathcal{A}'$ for any bounded Borel set $\mathcal{A}'\subset\mathbb{R}^{\mathrm{d}'}$ as the probability that $\hat{\Phi}(\mathcal{A}')$ is equal to zero where $\hat{\Phi}(\mathcal{A}')$ denotes the number of nodes in $\hat{\Phi}$ enclosed in $\mathcal{A}'$. That is,
$$\mathbb{P}\left[\hat{\Phi}(\mathcal{A}')=0\right] =\exp\left(-\int_{\mathcal{A}'} \hat{\Lambda}(\dif X')\right)$$
because $\hat{\Phi}$ is a PPP. Let $\mathcal{A}_i=\mathbf{T}_i(\mathcal{A}')$ where $\mathcal{A}_i\subset\mathbb{R}^{\mathrm{d}}$ is a bounded Borel set and the probability that node $X^{\dag}_i$ in $\hat{\Phi}^{\dag}$ is enclosed in $\mathcal{A}_i$ is
\begin{align*}
\mathbb{P}\left[X_i^{\dag}\in\mathcal{A}_i\right]&=\mathbb{P}\left[\mathbf{T}_i(X_i)\in\mathcal{A}_i \right]= \mathbb{P}\left[\mathbf{T}^{\mathrm{T}}_i\mathbf{T}_i(X_i)\in \mathbf{T}^{\mathrm{T}}_i(\mathcal{A}_i) \right]  \\
&= \mathbb{P}\left[X_i \in(\mathbf{T}^{\mathrm{T}}_i\mathbf{T}_i)^{-1}(\mathcal{A}') \right],
\end{align*}
where $(\mathbf{T}_i^{\mathrm{T}}\mathbf{T}_i)^{-1}(\mathcal{A}') \defn \int_{\mathcal{A}'}(\mathbf{T}_i^{\mathrm{T}}\mathbf{T}_i)^{-1}(\dif X)$.  By definition, $\hat{\Lambda}^{\dag}(\mathcal{A})$ is the average of $\hat{\Phi}^{\dag}(\mathcal{A})$ and it can be calculated by applying the Campbell theorem as shown in the following:
\begin{align*}
\hat{\Lambda}^{\dag}(\mathcal{A})&=\int_{\mathcal{A}}\hat{\Lambda}^{\dag}(\dif X^{\dag}) =\mathbb{E}\left[\sum_{X^{\dag}_i\in\hat{\Phi}^{\dag}}\mathds{1}\left(X^{\dag}_i\in\hat{\Phi}^{\dag}\cap\mathcal{A}_i\right)\right]\\
&=\mathbb{E}\left[\sum_{X_i\in\hat{\Phi}}\mathds{1}\left(X_i\in\hat{\Phi}\cap (\mathbf{T}_i^{\mathrm{T}}\mathbf{T}_i)^{-1}(\mathcal{A}')\right)\right]\\
&=\mathbb{E}\left[ \int_{ (\mathbf{T}^{\mathrm{T}}\mathbf{T})^{-1}(\mathcal{A}') }\mathbb{P}\left[X\in\hat{\Phi} \right] \nu_{\mathrm{d}'}(\dif X)\right]\\ &\stackrel{(a)}{=}\mathbb{E}\left[\frac{1}{\sqrt{\det(\mathbf{T}^{\mathrm{T}}\mathbf{T})}}\right]\int_{\mathbb{R}^{\mathrm{d}'}}\mathbb{P}\left[X\in\hat{\Phi}\cap\mathcal{A}' \right] \nu_{\mathrm{d}}(\dif X')\\
&=\mathbb{E}\left[\frac{1}{\sqrt{\det(\mathbf{T}^{\mathrm{T}}\mathbf{T})}}\right] \int_{\mathcal{A}'} \hat{\Lambda}(\dif X')
\end{align*}
where $(a)$ follows from the Jacobian determinant of two volumes. Thus,  $\hat{\Lambda}^{\dag}(\mathcal{A})=\mathbb{E}\left[\frac{1}{\sqrt{\det(\mathbf{T}^{\mathrm{T}}\mathbf{T})}}\right]\hat{\Lambda}(\mathcal{A}')$ and it follows that
\begin{align*}
\mathbb{P}\left[\hat{\Phi}(\mathcal{A}')=0\right]&=\exp\left(-\mathbb{E}\left[\frac{1}{\sqrt{\det(\mathbf{T}^{\mathrm{T}}\mathbf{T})}}\right]^{-1}\int_{\mathcal{A}'}\hat{\Lambda}^{\dag}(\dif X^{\dag})\right)\\
&=\mathbb{P}\left[\hat{\Phi}^{\dag}(\mathcal{A})=0\right]
\end{align*}
since $\nu_{\mathrm{d}}(\mathcal{A})=\nu_{\mathrm{d}'}(\mathcal{A}')$. Since the void probability of a point process completely characterizes the statistics of the process, $\hat{\Phi}^{\dag}$ is also a PPP.

 If $\hat{\Phi}$ is homogeneous, then its intensity measure is $\hat{\Lambda} (\mathcal{A}')=\hat{\lambda}\nu_{\mathrm{d}'}(\mathcal{A}')$ and thus the result in above can be further simplified to
 $\hat{\Lambda}^{\dag}(\mathcal{A})=\hat{\lambda}\mathbb{E}\left[ \sqrt{\det((\mathbf{T}^{\mathrm{T}}\mathbf{T})^{-1})}\right]\nu_{\mathrm{d}}(\mathcal{A})$ and it does not depend on the location of any node in $\hat{\Phi}^{\dag}$. As a result, $\hat{\Phi}^{\dag}$ is a homogeneous PPP with the intensity given in \eqref{Eqn:TransHomoIntensityMeasure}. 

\subsection{Proof of Theorem \ref{Thm:BoundVoidProb}}\label{App:ProofBoundVoidProb} 
 Suppose each user in cell $\mathcal{C}_l\subset\mathbb{R}^2$ of BS $B_l$ has its own cell association region which is also a bounded Borel set on $\mathbb{R}^2$. For example, the cell association region of user $U_j$ is denoted by $\mathcal{A}_j$ and all cell association regions contain  BS $B_l$. Since all users in $\mathcal{C}_i$ adopt the RCA scheme, the probability that BS $B_l$ is not associated by user $U_j\in\mathcal{C}_l$ can be expressed as
 \begin{align*}
 \mathbb{P}[\mathcal{E}_j]=&\lim_{\nu(\mathcal{A}_j)\rightarrow\infty}\mathbb{P}\bigg[W_lH_l\|U_j-B_l\|^{-\alpha}\\
 &<\sup_{B_i\in\Phi_{\subB}\cap\mathcal{A}_j\setminus B_l}\left\{W_iH_i\|U_j-B_i\|^{-\alpha}\right\}\bigg],
 \end{align*}
 where $\mathcal{E}_j$ denotes the event that user $U_j$ is not associated with BS $B_l$ as its association region goes to infinity. According to Theorem \ref{Thm:RanTransProp}, $\Phi_{\subB}$ can be statistically mapped to another homogeneous PPP $\Phi^{\dag}_{\subB}$ of intensity $\lambda^{\dag}_{\subB}=\lambda_{\subB}\mathbb{E}[(WH)^{\frac{2}{\alpha}}]$ and $\Phi^{\dag}_{\subB}(\mathcal{A}_j)$ is a Poisson random variable with parameter $\lambda^{\dag}_{\subB}\nu(\mathcal{A}_j)$. Thus, it follows that                      
 \begin{align*}
\mathbb{P}\left[\mathcal{E}_j\right]= \lim_{\nu(\mathcal{A}_j)\rightarrow\infty}&\mathbb{P}\bigg[\|U^{\dagger}_j-B^{\dag}_l\|^{-\alpha}<\sup_{B^{\dag}_i\in\Phi^{\dag}_{\subB}\cap\mathcal{A}_j\setminus B^{\dag}_l}\\
&\left\{\|U^{\dagger}_j-B^{\dag}_i\|^{-\alpha}\right\}\bigg]= \mathbb{P}\left[ D^{\dag}_j> D^{\dag *}_j\right],
 \end{align*}
 where $U^{\dagger}_j=(W_lH_l)^{-\frac{1}{\alpha}}U_j$ , $D^{\dag}_j$ is the distance from user $U^{\dagger}_j$ to BS $B^{\dag}_l$,  $D^{\dag *}_j$ is the distance from  $U_j$ to its nearest BS in $\Phi^{\dag}_{\subB}$.  The void probability $p_{\emptyset}$ can be found by calculating the following
$$p_{\emptyset}=\mathbb{E}\left\{\mathbb{P}\left[\bigcap_{j=0}^{\Phi_{\subU}(\mathcal{C}_l)} \mathcal{E}_j\bigg| \Phi_{\subU}(\mathcal{C}_l)\right]\right\}$$
Due to the complicate correlation effects between all events $\mathcal{E}_j$'s, it is analytically intractable to solve the exact result of $p_{\emptyset}$. However, the bounds on $p_{\emptyset}$ can be characterized in closed-form. 

First, the lower bound on $p_{\emptyset}$ can be found as follows. For a given $\Phi_{\subU}(\mathcal{C}_l)$ we know the following inequality
$$\mathbb{P}\left[\bigcap_{j=0}^{\Phi_{\subU}(\mathcal{C}_l)} \mathcal{E}_j\bigg|\Phi_{\subU}(\mathcal{C}_l) \right]\geq \prod_{j=0}^{\Phi_{\subU}(\mathcal{C}_l)}\mathbb{P}[\mathcal{E}_j],$$
which is obtained by considering $\mathbb{P}[\mathcal{E}_j|\mathcal{E}_i]\geq\mathbb{P}[\mathcal{E}_j]$ for $\mathcal{A}_i\cap\mathcal{A}_j\neq\emptyset$.  Since the number of users in $\mathcal{C}_l$ is a Poisson random variable with parameter $\lambda_{\subU}\nu(\mathcal{C}_l)$, i.e., $\mathbb{P}[\Phi_{\subU}(\mathcal{C}_l)=n]=\frac{(\lambda_{\subU}\nu(\mathcal{C}_l))^n}{n!}e^{-\lambda_{\subU}\nu(\mathcal{C}_l)}$,  the lower bound on $p_{\emptyset}$ can be given by
\begin{align}
p_{\emptyset} \geq  \mathbb{E}\left\{ \prod_{j=0}^{\Phi_{\subU}(\mathcal{C}_l)}\mathbb{P}[\mathcal{E}_j]\right\}
= \sum_{n=0}^{\infty}\left(\prod_{j=0}^{n}\mathbb{P}[\mathcal{E}_j]\right) \frac{(\lambda_{\subU}\nu(\mathcal{C}_l))^n}{e^{\lambda_{\subU}\nu(\mathcal{C}_l)}n!}. \label{Eqn:LowBoundVoidProb02}
\end{align}
According to the Slivnyak theorem \cite{DSWKJM96}, the statistic property evaluated at any point in a homogeneous PPP is the same such that  $\mathbb{P}[\mathcal{E}_j]$ is the same as $\mathbb{P}[\mathcal{E}_0]$ evaluated at user $U_0$ (the origin) for all $j$. Accordingly, \eqref{Eqn:LowBoundVoidProb02} can be simplified as
\begin{equation}
p_{\emptyset} \geq   \exp\left(-\lambda_{\subU}\nu(\mathcal{C}_l)(1-\mathbb{P}[\mathcal{E}_0])\right) \label{Eqn:LowBoundVoidProb03}
\end{equation}
and $\mathbb{P}\left[\mathcal{E}_0\right]=\mathbb{P}\left[ D^{\dag}_0> D^{\dag *}_{0}\right]
$ in which $D^{\dag *}_0$ is the distance from the origin to the nearest BS in $\Phi^{\dag}_{\subB}$. The pdf of $D_0^{\dag *}$ is $f_{D_0^{\dag *}}(x)=2\pi\zeta^{\dag}\lambda_{\subB}x\exp(-\pi\zeta^{\dag}\lambda_{\subB}x^2)$ \cite{FBBBL10}, whereby $\mathbb{P}[\mathcal{E}_0]$ is simplified as
 \begin{align}
 \mathbb{P}[\mathcal{E}_0] =1- \mathbb{E}\left[\exp\left(-\pi\zeta^{\dag}\lambda_{\subB}(D_0^{\dag })^2\right)\right].
 \end{align}   
Since all users in $\mathcal{C}_l$ are uniformly distributed, we can know $f_{D^{\dag}_0}(r)=\frac{2\pi r\zeta^{\dagger}}{\nu(\mathcal{C}_l)}$ and thus we can calculate $\mathbb{P}[\mathcal{E}_0]$ in closed-form as
\begin{align*}
\mathbb{P}[\mathcal{E}_0] &= 1-\int_{0}^{\sqrt{\frac{\nu(\mathcal{C}_l)}{\pi}}} \exp\left(-\pi\zeta^{\dag}\lambda_{\subB}r^2\right)\frac{2\pi r\zeta^{\dagger} }{\nu(\mathcal{C}_l)}\dif r\\
 &= 1-\frac{1-\exp(-\zeta^{\dagger}\lambda_{\subB}\nu(\mathcal{C}_l))}{\lambda_{\subB}\nu(\mathcal{C}_l)}.
\end{align*}
Apparently, $\mathbb{P}[\mathcal{E}_0] \geq 1-\frac{1}{\lambda_{\subB}\nu(\mathcal{C}_l)}$. Then substituting it into \eqref{Eqn:LowBoundVoidProb03} gives us the lower bound in \eqref{Eqn:BoundVoidProb}. 

Next, the upper bound on $p_{\emptyset}$ can be obtained as follows. The probability that  user $U_j$ does not associate with BS $B_l$ is upper-bounded by
\begin{align*}
 \mathbb{P}[\mathcal{E}_j]\leq\lim_{\nu(\mathcal{A}_j)\rightarrow\infty}& \mathbb{P}\bigg[W_lH_l\inf_{U_j\in\mathcal{C}_l}\{\|U_j-B_l\|^{-\alpha}\}< \\ &\sup_{B_i\in\Phi_{\subB}\cap\mathcal{A}_j\setminus B_l}W_iH_i\|U_j-B_i\|^{-\alpha}\bigg],
\end{align*}
which is acquired by considering the farthest user of BS $B_l$ in cell $\mathcal{C}_l$.  According to Slivnyak's theorem and Theorem \ref{Thm:RanTransProp}, the upper bound on $p_{\emptyset}$ for a given  $\Phi_{\subU}(\mathcal{C}_l)$  can be written as
$$\mathbb{P}\left[\bigcap_{j=0}^{\Phi_{\subU}(\mathcal{C}_l)} \mathcal{E}_j\bigg|\Phi_{\subU}(\mathcal{C}_l) \right]\leq \left(\mathbb{P}\left[R^{\dag}_l \geq D^{\dag *}_0\right]\right)^{\Phi_{\subU}(\mathcal{C}_l)},$$
where $R^{\dag}_l$ denotes the distance from BS $B^{\dag}_l$ to its farthest user in $\mathcal{C}^{\dagger}_l=(WH)^{-\frac{1}{\alpha}}\mathcal{C}_l$. Let $\Phi^{\dagger}_{\subU}(\mathcal{C}^{\dagger}_l)\defn\{U_j^{\dagger}\in\mathcal{C}^{\dagger}_l: (W_jH_j)^{-\frac{1}{\alpha}}U_j, U_j\in\mathcal{C}_l\}$  and for given $\Phi_{\subU}(\mathcal{C}_l)$  we have
\begin{align*}
&\mathbb{P}\left[R^{\dag}_l \geq D^{\dag *}_0\right]=\mathbb{P}\left[(R^{\dag}_l)^2 \geq (D^{\dag *}_0)^2\right]\\
&\stackrel{(a)}{=}\mathbb{P}\left[(W_lH_l)^{-\frac{2}{\alpha}}\sum^{\Phi^{\dagger}_{\subU}(\mathcal{C}^{\dagger}_l)}_{j=1}R_{0,j}^2 \geq (D^{\dag *}_0)^2\right]\\
&\leq \mathbb{E}\left\{\prod_{j=1}^{\Phi^{\dagger}_{\subU}(\mathcal{C}^{\dagger}_l)}\mathbb{P}\left[(W_lH_l)^{-\frac{2}{\alpha}}R_{0,j}^2 \geq (D^{\dag *}_0)^2\right]\right\}\\
&= \left(1-\exp\left(-\pi\lambda^{\dagger}_{\subB} \mathbb{E}\left[(W_lH_l)^{-\frac{2}{\alpha}}\right]R_0^2\right)\right)^{\mathbb{E}[\Phi^{\dagger}_{\subU}(\mathcal{C}^{\dagger}_l)]}\\
&\stackrel{(b)}{\leq}  \left(1-\frac{\lambda_{\subU}}{\lambda_{\subU}+\mathbb{E}[(WH)^{-\frac{2}{\alpha}}]\lambda^{\dagger}_{\subB}}\right)^{\Phi_{\subU}(\mathcal{C}_l)/\zeta^{\dagger}}\\
&\stackrel{(c)}{\leq}  \left(1+\frac{\lambda_{\subU}}{\zeta^{\dagger}\lambda_{\subB}}\right)^{-\zeta^{\dagger}/\Phi_{\subU}(\mathcal{C}_l)}
\end{align*}
where (a) follows from that fact that $\{(R^{\dagger}_l)^2, l=1,2,\cdots\}$ is an one-dimensional Poisson process of intensity $\lambda_{\subU}\mathbb{E}[(WH)^{\frac{2}{\alpha}}]$ and thus $(R^{\dagger}_l)^2$ is equal to the sum of $\Phi_{\subU}(\mathcal{C}_l)$ i.i.d. $(WH)^{-\frac{2}{\alpha}}R^2_0$ \cite{MH05}, $(b)$ is due to first applying Jensen's inequality on the term $(WH)^{-\frac{2}{\alpha}}$ and then calculating the expectation regarding $R_0^2$, and $(c)$ is obtained from $(1-x)^n\leq (1-x)^{1/n}$ for $x\in[0,1]$ and $n>0$ and $\mathbb{E}[\Phi_{\subU}(\mathcal{C}^{\dagger}_l)]=\Phi_{\subU}(\mathcal{C}_l)/\zeta^{\dagger}$. Therefore, we have
$$\mathbb{P}\left[\bigcap_{j=0}^{\Phi_{\subU}(\mathcal{C}_l)} \mathcal{E}_j\bigg|\Phi_{\subU}(\mathcal{C}_l) \right]\leq\left(1+\frac{\lambda_{\subU}}{\zeta^{\dagger}\lambda_{\subB}}\right)^{-\zeta^{\dagger}}$$
for any given $\Phi_{\subU}(\mathcal{C}_l)$ and the upper bound in \eqref{Eqn:BoundVoidProb} follows. This completes the proof. 

% section of references

\bibliographystyle{ieeetran}
\bibliography{IEEEabrv,Ref_CellAssVoidProb}

\end{document}